\documentclass[prb,preprint]{revtex4-1} 
\usepackage{amsmath} 
\usepackage{amsfonts} 
\usepackage{graphicx} 
\begin{document}
\title{Friction of spheres on a rotating parabolic support} 
\author{Alexis Soulier}
\affiliation{SPHYNX/SPEC, CEA-Saclay, Orme les merisiers, F-91191 Gif-sur-Yvette, France }
\author{S\'ebastien Auma\^{\i}tre}
\email{sebastien.aumaitre@cea.fr}
\affiliation{SPHYNX/SPEC, CEA-Saclay, Orme les merisiers, F-91191 Gif-sur-Yvette, France }
\date{\today}
\begin{abstract}
This article illustrates the role of friction on the motion of a rolling sphere on pedagogical example. We use a parabolic support rotating around it axis to study the static equilibrium positions of a single sphere. Due to the particular choice of the shape of support, some easy analytical calculations allow theoretical predictions. (i) In the frictionless case, there is an eigen frequency of rotation where the gravity balances the centrifugal force. All positions on the parabola are therefore in static equilibrium. At others rates of rotation, the sphere can go to the center or escape to infinity. It depends only on the sign of the detuning with the eigenfrequency. (ii) In contrast, we show that the static friction imposes a range of equilibrium positions at all rotating rates. These predictions can be compared to the maximum equilibrium radius measured on the experimental device. A reasonable estimate of the static friction between the support and spheres made of different materials can be extracted from this maximum equilibrium radius. To go further in understanding of the experimental results, we perform a stability analysis also in the case with friction. This analysis involves the rolling friction. We show that this rolling friction can be estimated in the same device just by the check the dissipation during the motion of the sphere.
\end{abstract}
\maketitle 
\section{Introduction} 
\begin{figure}
\centering
\includegraphics[width= .80\columnwidth,angle=0]{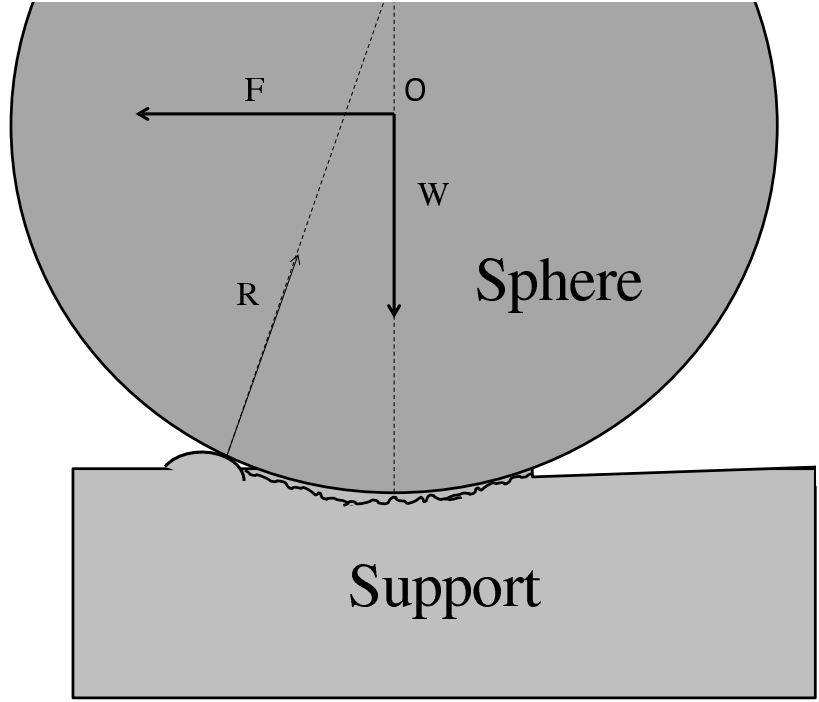}
\caption{Sketch of the contact of a rolling sphere of weight {\bf $W$} pulled with a tangential force $F$ on a support. {\bf $R$} is the reaction of the support resulting from the contact points between sphere and support.}
\label{contact}
\end{figure}

The friction is a puzzling non-conservative contact force\cite{FeymannI}. Indeed, the relative motion of block parallel to a surface has two aspects. First, one has to apply a finite force tangential to surface to overcome the {\it static} friction and set the block into motion. Once the block is in motion, the {\it sliding} friction resists to the displacement and induces some dissipation. In both cases, these frictional forces are proportional to the load applied on the block. However the coefficients of proportionality, called {\it friction coefficients}, are different. The static friction coefficient is usually the larger. These different friction coefficients imply a lot of nonlinear behaviors. Actually, it contributes to a great part of the richness of the physics of granular packing. Moreover they are involved in natural phenomena such as earthquakes, avalanches, arches effects, grains segregation etc. ....\cite{Bideau}
Friction is even more puzzling when the spheres are concerned. Indeed, they imply three types of frictions. A perfectly non deformable sphere will roll on a perfectly flat support without dissipation as soon as a tangential force is applied. This is due to the immobility of the single contact point during the rolling. However a real sphere is slightly deformed by, and/or slightly deforms, its support as emphasized figure \ref{contact}. Therefore the contact is made through few points due to rugosity of the surfaces in contact \cite{LandauI,Witters}. Hence, the reaction of the support is, neither exactly parallel to the weight nor exactly aligned with the center of sphere as shown in figure \ref{contact}. Therefore, a minimum torque has to be applied to induce the rolling. This induces a {\it static friction} before the rolling of a real sphere. Moreover, above this threshold, the rolling involves rejuvenation of the contact points. This process implies some energy lost by visco--plastic deformations, by hysteretic elastic deformations, by adhesion, by electrostatic interactions, etc... \cite{Poschel,Johnson}. Therefore, a {\it rolling friction} is produced at the surface of contact, even if there is no relative tangential velocity between the sphere and the support at the contact. The {\it sliding friction} comes next when there is a relative speed at the contact \cite{Domenech2,Shaw}.
The rolling motion of a sphere involves the balance of forces and torques. However, it can be shown that the problem reduces to a balance of force acting on the center of mass of a sphere of uniform density \ cite {Witters, Kondic}. Due to the rolling, the static frictions of spheres and rods are generally much lower than for other sliding bodies. Indeed, a very small slope is usually enough to put a sphere into a rolling motion. The measurement of static friction by such methods is quite difficult. In addition, the static friction is subject to aging and can be sensitive to the local roughness, so the exact value depends on the measurement protocol.
There are many mechanisms involved in the rolling motion of a real sphere. Hence, it seems useful to offer at the students of a basic lecture of mechanics some pedagogical examples where measurement can be compared to tractable theoretical predictions.
Here, we propose a simple experiment using a centrifuged sphere on a parabolic support. On this device, for instance, we show that the particle, starting from an out-of-equilibrium position, stops at a given equilibrium radius, $r _{eq}$ of the parabola. This equilibrium radius tends to increase with the rotation frequency. One can compute the upper bound of $r_{eq}$ at a given rotation rate. It is parameterized by the static friction. Therefore the curve of the maximum of $r_{eq}$ versus the rotation frequency $\Omega$ gives an estimate of the static friction coefficient.
In the first part, we describe the experimental setup and the measurement techniques. Then we perform the force balance and we compute the static positions accessible for the sphere. These theoretical predictions are compared to the experimental measurements. To go further in the comparison, a stability analysis of the static positions is also discussed. This analysis involves the rolling friction. Therefore we study the transient regime before the static equilibrium. Indeed, the rolling friction can be extracted from the damped motion of the sphere to its equilibrium position. In the conclusion, we give some possible extensions of this work. 
\section{Experimental setup}
\begin{figure}[h!]
\centering
\includegraphics[width=5in]{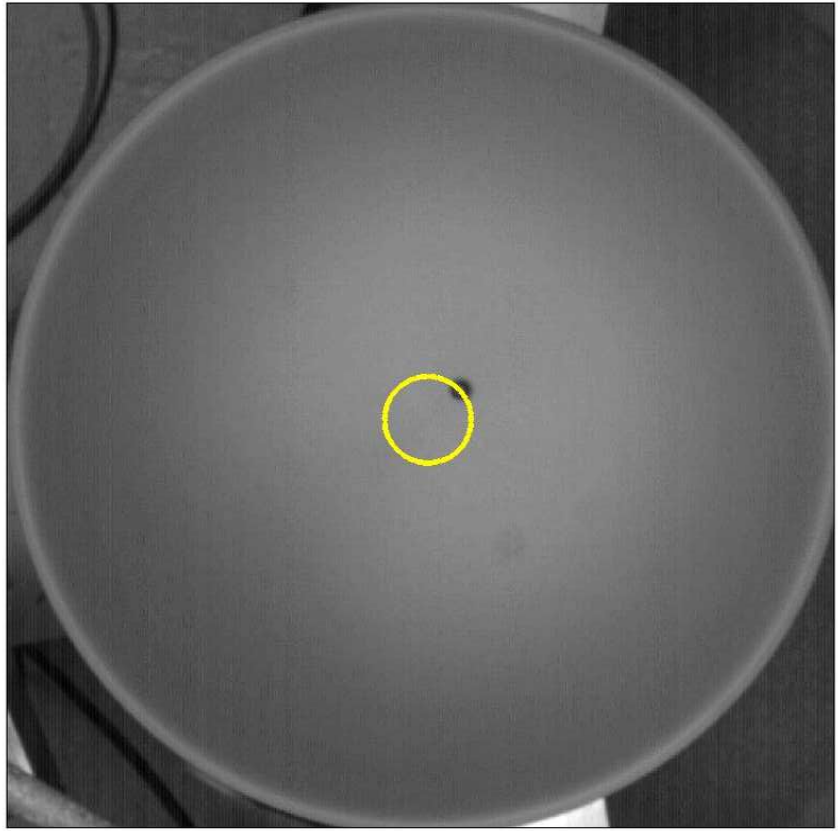}
\caption{Snapshot of the particle position in the stationary regime for a rotation rate of 6.28 radiant/s. Particle is a stainless Steel about 6 mm diameter. The yellow circle shows the tracked trajectory in the lab frame.}
\label{PosTrac}
\end{figure}

The setup, sketched in figure \ref{Fig DevCent}--a, consists on a parabolic vessel. The parabola has a maximum diameter of 250 mm for a maximum depth of 35 mm. Hence, its shape is described by a function $Z(r)=ar^2$ along the vertical axis, with $a=2.24\times 10^{-3}$mm$^{-1} $. The parabola is machined with a precision of XX. Its surface is made in anodized aluminum with a polishing about 1 $\mu$m. The vessel is driven into rotation around the vertical axis at a constant frequency by a step motor. A single sphere put on the parabolic support. Different spherical materials have been tested: stainless steel spheres of a ball bearing with a diameter of 6 and 3 mm and a glass sphere of 6 mm diameter have been used. The Tab.~1 gives the characteristics of these spheres. Two particles (named P1 and P2) are almost identical stainless steel spheres, P3 is made of glass, P4 is a twice smaller stainless steel sphere. The motion of the sphere on the parabola is followed from above with a CMOS camera of 1280X1024 pixels. The particle resolution is about 24 pixels per particle diameter for the spheres of 6 mm.
\begin{tabular}{|c| c| c| c | c| p{3.5cm}|}
\hline 
Particles& $d$ (mm) & Material &Masse (g) & Rugosity ($\mu$m)&Deviation from
spherical form($\mu$m)\\
\hline 
P1 & 6.2 & Stainless Steel& 1.06&$0.014$&$0.13$\\
P2 & 6.1&Stainless Steel &1.06&$0.014$&$0.13$\\ 
P3 & 6.0 & Glass &0.285&$0.15$&$5$\\
P4 & 3.0&Stainless Steel &0.106&$0.01$&$0.08$\\
\hline
\end{tabular}
\noindent
\centerline{Tab. 1. Charactersitic of the four spheres used in the experiments\\}
\begin{figure}[h!] 
\centering
\includegraphics[width= 0.99\columnwidth,angle=0]{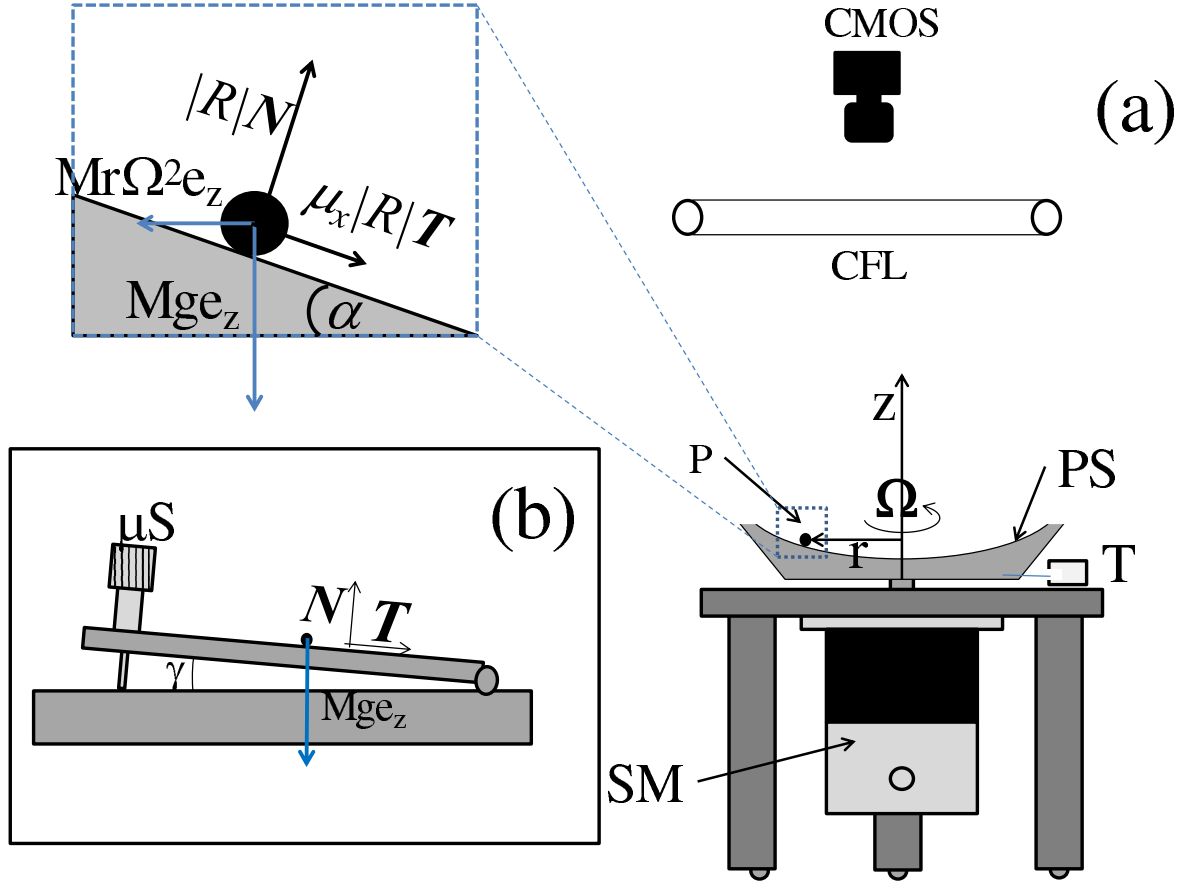}
\caption{Experimental devices: (a) The Step Motor (SM) drives the parabolic Support (PS) at the angular velocity $\Omega$ around the $z$-axis. A spherical particle (P) put on the parabolic support is tracked CMOS camera. All the device is enlighten by a Circular Fluorescent Light (CFL). The tachometer (T) allows the control of the step motor and a stroboscopic image acquisition. The inset shows the force balance with $\mathbf{ R}$ the support reaction to the gravity $Mg\mathbf{e_z}$ and the centrifugal force, $Mr\Omega^2\mathbf{e_r}$ and $\mu \mathbf{R}$ the friction force. The angle $\alpha$ is the local slope of the parabola i.e.
$\tan(\alpha)=dZ/dr$ with $Z(r)=ar^2$ the shape of the support. (b) The inclined plan using a micrometric screw ($\mu$S) to measure the critical angle $\gamma_c$ above which the sphere starts rolling.} 
\label{FigDevCent}
\end{figure}
The experimental procedure is the following.
We launch the particle at an out-of-equilibrium position at the external radius of the parabola. For each particles several initial conditions are probed for a given rotation rate. We realized the experimental runs for a given particle at different days. Then the sphere is tracked up to its final position, static in the rotating frame. Acquisition can be started as soon as the sphere is launched. The images acquisition rates can go up to 25 frames/s. In order to follow the particle motions in the co-moving frame, the acquisition rate can also be triggered by the rotation frequency of the parabola, by the use of homemade optical tachymeter. Up to 600 pictures are taken in this case. However, for most of the measurement presented hereafter, the acquisition is started once the particle seems at rest in the rotating frame. This state is pointed out by the fact that the sphere does not emit the sounds induced by the rolling anymore. From 100 to 300 stroboscopic images are then captured.
We enlighten the parabola with a circular fluorescent light. In order to remove the light frequencies matching with the acquisition rate of the images, the fluorescent light is driven only at high frequency. This circular lamp makes a clear bright spot on the top of the sphere. This spot can be tracked by the use of any image analysis software. Figure \ref{PosTrac} illustrates the tracking efficiency for a stainless steel sphere of 6 mm. The circular yellow line plotted on the figure \ref{PosTrac}, shows the equilibrium position of particle in the lab frame at a rotation rate of $\Omega=6.28 $ rad/s.
In addition to this device, we built an aluminum inclined plane with the same anodized surface than the parabola. We use it to determine the angle of repose of the spheres under study and thus, their static friction. This angle of repose is the one below which the sphere stays 
at rest on the inclined plan and above which it starts rolling A simple force balance shows that this angle gives a direct access to the static friction. Indeed, following the sketch in inset figure \ref{FigDevCent}--b, a sphere starts to roll as soon as $Mg\mathbf{T}\mathbf{e_z}\geq\mu_o Mg\mathbf{N}\mathbf{e_z}$ where $M$ is the mass of the sphere, $g$ is the gravitational acceleration, $\mathbf{e_z}$ is the vertical unit vector pointing down, $T$ (respectively $N$) is the unit vector parallel (resp. perpendicular) to the inclined plane . Since $\mathbf{N}\mathbf{e_z}=\cos(\gamma)$ and $\mathbf{N}\mathbf{e_z}=\cos(\gamma)$, one has the relation $\mu_o^*=tan(\gamma_c)$ between the angle of repose, $\gamma_c$ and the static friction of the sphere measured on the inclined plane $\mu_o^*$. We used the following procedure to determine the values reported in Tab. 2. We start with the sphere at rest on the horizontal plane. Then we increase the angle $\gamma$ with a micrometric screw. $\gamma_c$ is the angle at which the particle rolls down up to bottom the inclined plane. We did not take into account the small motions of the sphere that occur sometime before the rolling of the sphere up to the bottom of the slope. Hence, we may overestimate the friction. About 21-28 measurements are performed for a single particle. Each time, the initial starting point is changed. The high level of the variability of this measurement has to be noted. The error, based on the standard deviation at the 21 measurements, is about 33\% (it would have been about 100\% based on the extreme values). These variations are mainly due to the change of the initial contact point at each measurement. In some experiments, the particle was guided in semispherical gutter or a square groove. That does change neither the mean value nor the standard deviation.
\begin{center}
\begin{tabular}{|c|p{3.5cm}| p{3.5cm} | p{3.5cm}|}
\hline 
Particles& Rolling angle $\gamma_c^*$&Static friction $\mu_s^*$& Rolling friction $\mu_r$\\
\hline 
P1 &$1.7\pm0.7$&$0.03\pm0.01$&$4.9\pm0.12 \times 10^{-4}$\\
P2 &$2.0\pm0.7$&$0.035\pm0;01$&$5.5\pm0.12 \times 10^{-4}$\\ 
P3&$1.7\pm0.6$&$0.03\pm0.01$&$4.9\pm0.14 \times 10^{-4}$\\
P4 &$0.04\pm0.015$&$2.4\pm0.7$&$6.9\pm0.5 \times 10^{-4}$\\
\hline
\end{tabular}
\medskip
\noindent
Tab. 2. Physical properties of the spheres, measured with the inclined plan experiments for rolling angle and the static friction, measure on the parabola for the rolling friction\\
\end{center}
\section{Static equilibrium}
Qualitatively, we observed on the parabola that the particle reaches an equilibrium position $r_{eq}$ at a small rotation rate $\Omega$. This value is nearly zero for the smallest rotation rates but increases with $\Omega$. Above an onset $\Omega_o\sim$6.63 rad/s, we note that the particle always escapes from the parabola. The value of the onset seems to almost not depend on the sphere material and diameter. In order to understand these qualitative observations, in this section, we shall consider the forces acting on the center of mass of sphere of mass. These forces are:
\begin{itemize}
\item the gravity force, $ -Mg\mathbf{e_z}$ acting downward along the vertical axis;
\item the centrifugal force $ M\Omega^2r\mathbf{e_r}$ acting on the radial direction $\mathbf{e_r}=\mathbf{r}/|r|$; 
\item the Coriolis force $- M\Omega\dot{r}\mathbf{e_\theta}$ acting in the direction $\mathbf{e_\theta}$ perpendicular to $\mathbf{e_r}$ and $\mathbf{e_z}$;
\item the reaction of the support $\mathbf{R}$ to the centrifugal and gravity force, acting perpendicularly to the surface of the parabola; 
\item the friction force $\mu_x R \mathbf{T}$ proportional to the force normal to the surface, acting in the direction of the unit vector $\mathbf{T}$ parallel to the surface of the parabola. 
\end{itemize}
They are shown in the inset of the figure \ref{FigDevCent} (except the Coriolis force which acts perpendicularly to the plan of the figure).
When no motion occurs, the static friction, $\mu_o R$, gives the onset that any forces applied tangentially to the surface have to overcome in order to induce a motion. Hence we do not see any motion as long as:
\begin{equation}
M |\Omega^2r \mathbf{e_r}\cdot\mathbf{T}- g\mathbf{e_z}\cdot\mathbf{T}|\leq \mu_o R
\label{StatEq}
\end{equation}
\noindent
Considering figure \ref{FigDevCent}, one finds that $\mathbf{e_r}\cdot\mathbf{T}=\cos(\alpha)$ and $\mathbf{e_z}\cdot\mathbf{T}=\sin(\alpha)$. The reaction of the support induced by the gravity and the centrifugal force is:
\begin{eqnarray}
R=&M( \Omega^2r \mathbf{e_r}\cdot\mathbf{N}- g\mathbf{e_z}\cdot\mathbf{N})\\ \nonumber
R=&M\left(\Omega^2r \sin(\alpha)+g\cos(\alpha)\right)
\label{Reaction}
\end{eqnarray}
\noindent
where $\mathbf{N}$ is the unit vector perpendicular to the parabolic surface. Moreover following figure \ref{FigDevCent}, one has $\mathbf{e_r}\cdot\mathbf{N}=-\sin(\alpha)$ and $\mathbf{e_z}\cdot\mathbf{N}=\cos(\alpha)$. In the specific case of a parabolic support, $\tan(\alpha)$ reduce to $\tan(\alpha)=2ar$.
\subsection{The frictionless case $\mu_o=0$}
When $\mu_o=0$, equation (\ref{StatEq}) simplifies to 
\begin{equation}
(\Omega^2-\Omega_o^2)r_{eq}=0.
\label{frictionless}
\end{equation}
\noindent
$\Omega Q=\sqrt{2 ga}=6.63$ rad/s is the eigenfrequency of the rotating parabola. It corresponds to the frequency where the gravity and the centrifugal force acting on sphere, balance each other. At this rotation rate, all the positions on the parabola correspond to a static equilibrium for a frictionless sphere.
\subsection{The frictional case $\mu_o\neq 0$}
\begin{figure}
\centering
\includegraphics[width= 0.9\columnwidth,angle=0]{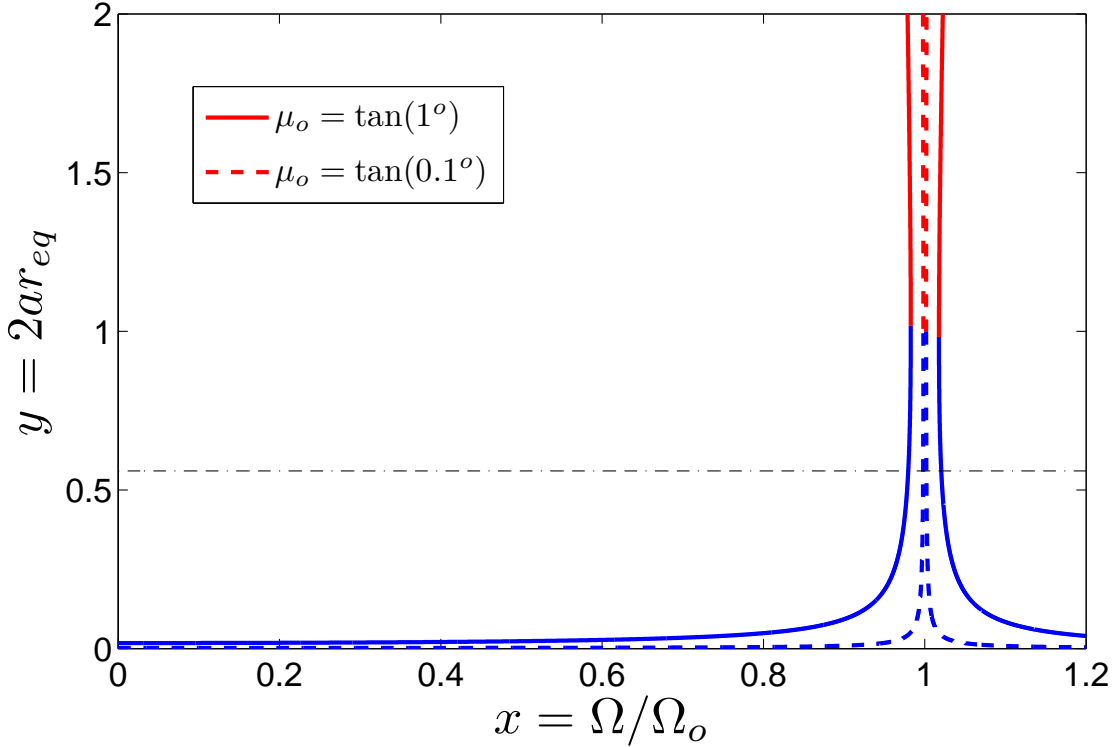}
\caption{Boundaries of the static positions of a frictional sphere on a rotating parabola. Red lines gives $y^+=2ar_{max}^+$ the largest root of equation (\ref{frictionpolyn}) whereas the blue lines represent $y^-=2ar_{max}^-$. The dashed line shows the roots for $\mu_o=tan(0.1^o)=0.0017$ the full line are computed for$\mu_o=tan(1^o)=0.0175$ . The horizontal dashed line shows cell size. }
\label{req}
\end{figure}

In this case, the static radius, $r_{eq}$, is constrained, from eq. (\ref{StatEq}), by the condition :
\begin{equation}
2a\mu_o \Omega^2 r_{eq}^2-|\Omega^2-\Omega_o^2|r_{eq}+\mu_o g\leq0. 
\label{frictionpolyn}
\end{equation}
\noindent
The maximum radius of the static position, $r_{max}$, is given by the roots of the left hand side of (\ref{frictionpolyn}):
\begin{equation}
y^\pm=\frac{|x^2-1|\pm\sqrt{|x^2-1|^2-4\mu_o^2 x^2}}{2\mu_o x^2}.
\label{rootpolyn}
\end{equation}
\noindent
We introduce here the dimensionless parameters $x=\Omega/\Omega_o$ and $y=2ar_{max}$. These roots have an imaginary part for: $|x^2-1|^2-4\mu_o^2 x^2<0$. They define an extended range of rotation rate $\tilde{\Omega}$ around $\Omega_o$, such that $|\tilde{\Omega}^2/\Omega_o^2-1|^2\Omega_o^2/(4\tilde{\Omega}^2)<\mu_o$, where all the radii satisfy the inequality (\ref{frictionpolyn}). Actually, it is a reminiscence of the frictionless case where this degeneracy occurs only when $\Omega=\Omega_o$ . For instance, with $\mu_o<<1$ and thus, $\tilde{\Omega}/\Omega_o\sim 1$, one has all the domain : $\Omega_o(1-\mu_o)\leq\tilde{\Omega}\leq \Omega_o(1+\mu_o)$ where all positions on the parabola correspond to a static equilibrium. From equation (\ref{rootpolyn}), one has to note that the range of possible static radii depends only on $\mu_o$, for a given setup. Thus, it could be a way to estimate the static frictions of a sphere. The domains of static positions allowed by equation (\ref{frictionpolyn}) are delimited on figure \ref{req} for two friction coefficients $\mu_o=0.0175$ (corresponding to an angle of repose on a plan inclined about an angle of $1^o$) and the other for $\mu_o=0.001=tan(0.1^o)$. In our finite size cell, $r_{max}^+=y^+/(2a)$ is always larger than the cell size as shown by the horizontal dashed line on figure \ref{req}. Hence, only the radius $r< r_{max}^-$ should be visible. Nevertheless, we must still check that all positions in these position ranges are stable against small perturbations. This is the purpose of the section III.D.
\subsection{Experimental results}
\begin{figure}
\centering
\includegraphics[width= 1.0\columnwidth,angle=0]{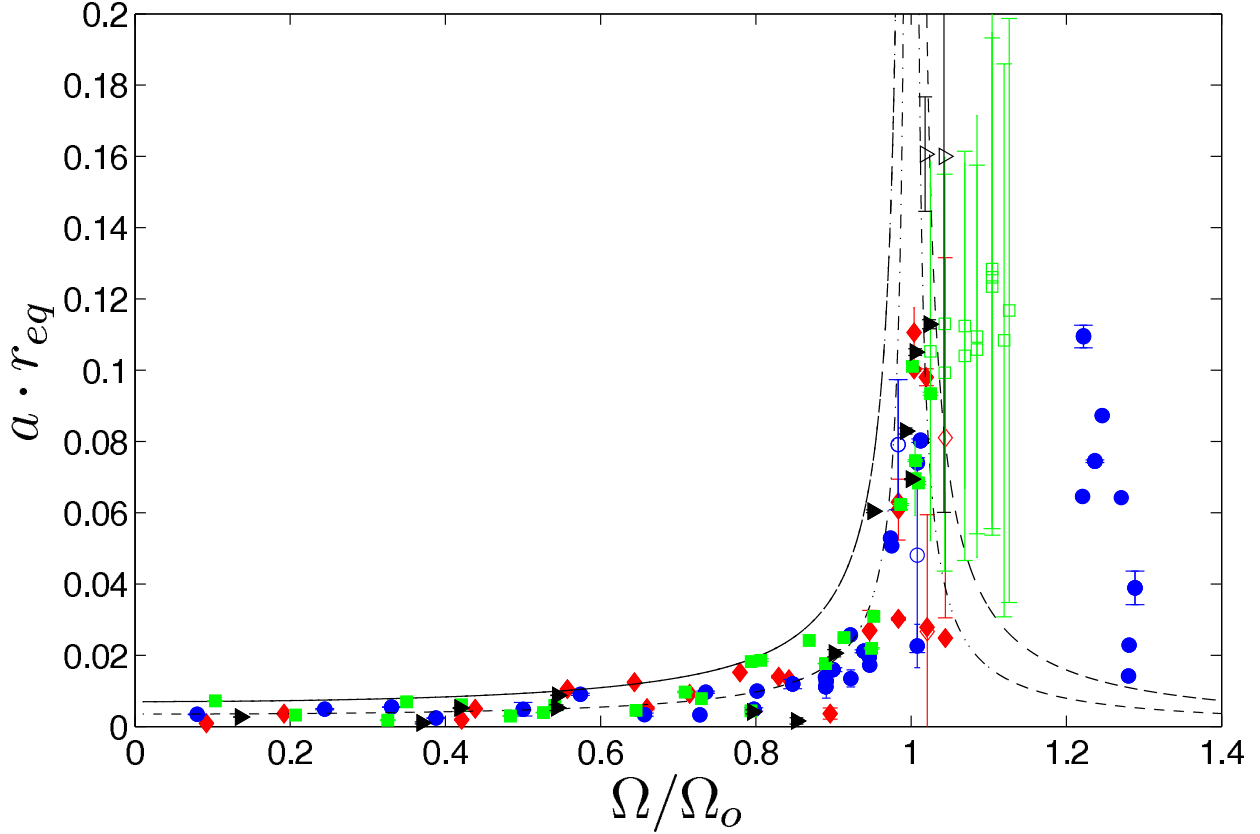}
\caption{Static radii of stainless steel sphere of 6 mm diameter (P1: red diamond) (P2: blue disk), glass sphere of 6 mm diameter (P3: green square) and Stainless steel sphere of 3 mm diameter (P4: black left triangle). The error bars are given by the particle displacement during the twenty last turns. The full symbols represent the stable static positions; the open symbols represent the unstable ones. The dashed line shows $y^-$ (equation (\ref{rootpolyn})) for $\mu_o=\tan(0.8^o)=0.014$. Dot--dashed line shows $y^-$ for $\mu_o=\tan(0.4^o)=0.007 $. }
\label{reqexp}
\end{figure}

We check in our experimental device the four particles resumed in Tab.\~1.
Figure \ref{reqexp} shows that most of the measured radius can be limited by the boundaries given by equation (\ref{rootpolyn}) using $\mu_o\sim \tan(0.8)=0.014$. In agreement with measurement on inclined plan, all particles have a close static friction. However the statics friction measured here seems to be more compatible with the lower bounds of the measurement on inclined plan (see Tab. 2). The smallest sphere, P4, reaches smaller radii especially at low rotation rate although it is made of the same material than the spheres P1 and P2. It could be due to its smaller size implying a lower mass and lower inertia. Hence the particle stops before exploring the all available domain. The small sphere could also be more sensitive to inhomogeneous asperities. Some puzzling measurements with a particle P2 are out of the range of acceptable measure (on the right of the figure \ref{reqexp}). It may be due to some (not visible) defects on the sphere surface that prevent the rolling, or to liquid bridges introduced during the cleaning procedure. 
Although we usually begin the measurement when the sphere appears motionless in the co-moving frame, it happens that these positions are not always stable. Sometime the particle starts to move up, first slowly then faster and faster, and finally escapes from the cell. These metastable radii are noted with open symbols in figure \ref{reqexp}. They appear with all the kinds of spheres and mostly for $\Omega/\Omega_o>1$. This is underlined on figure \ref{reqexp} by the larger error bars of these points. Indeed these error bars are based on the displacement during the 20 last turns. Note that this destabilization seems to occur also for rotation rate close to $\Omega_o$, therefore at a position probably included in the stationary cone defined by equation (\ref{rootpolyn}) whatever is the static friction. Moreover the time necessary before a significant motion of the particle can become quite long. Therefore, even if we have multiplied by three the measurement time for rotation rate near and above $\Omega=\Omega_o$, we cannot insure that all static positions measured, are in stable positions.
\subsection{Stability}
It seems therefore quite relevant to estimate the stability of the static equilibrium radius given by equation (\ref{rootpolyn}).
When friction is involved, the stability analysis becomes trickier because one has to deal with the discontinuity between the static and the rolling frictions. Hence, we will consider first the easier frictionless case although the result is intuitive. It will give us a guideline for the frictional case that we discuss just after.
\subsubsection{Frictionless case}
The question is here to understand where the particle moves when the rotation rate, $\Omega$, is slightly shifted from the resonant rotation rate $\Omega_o$. In order to check this, we restrict our study to the radial position of the sphere in cylindrical coordinates. From the fundamental dynamical principle, one can derive the following equation of motion:
\begin{equation}
\ddot{ r}= r\Omega^2 -\left( r\Omega^2 \sin(\alpha)+g\cos(\alpha)\right)\sin(\alpha) 
\label{Stability1}
\end{equation}
\noindent
where the first term is due to centrifugal force and the second is induced by the parabolic support reaction. We do not take into account the Coriolis force. Indeed, it is proportional to the particle velocity. Hence it is negligible regarding the study of the static equilibrium as long as the initial velocity perturbation is small. We Use first the trigonometric relation $\cos^2(\alpha)=1/(1+\tan^2(\alpha))$ and the fact that for a parabola, $\tan(\alpha)=2ar$. Then we multiply each sides of equation (\ref{Stability1}) by $\dot{r}$. It can be rewritten as:
\begin{equation}
\frac{1}{2}\frac {d \dot{ r}^2}{dt}= \frac{\left(\Omega^2 -\Omega_o^2 \right)r\dot{r}}{1+(2ar)^2} 
\label{Stability2}
\end{equation}
\noindent
This equation can be integrated to get
\begin{equation}
\frac {\dot{ r}^2}{2}=\delta E_o+ \left(\Omega^2 -\Omega_o^2 \right)\frac{\ln\left(1+(2ar)^2\right)}{8a^2} ,
\label{Stability3}
\end{equation}
\noindent
where $\delta E_o$ is an integration constant representing the small initial energy input necessary to disturb the position. In order to justify the removal of the Coriolis force, we must just assume that $\delta E_o$ implies an initial velocity perturbation $\sqrt{2 \delta E_o}<<r \Omega$ .
We see from the right hand side of the equations (\ref{Stability2}) that for $\Omega >\Omega_o$ the radial kinetic energy will increase and the particle will escape from the cell. The Coriolis force must be considered to get the escape time. For $\Omega <\Omega_o$ the radial kinetic energy will decrease except for $r=0$. Hence, the particle finds its full equilibrium at $r=0$. Note that the equation (\ref{Stability3}) can be seen as an energy budget $\delta E_o=K+V$, with $K= \dot{r}^2/2 $. The term $V(r)=-\left(\Omega^2 -\Omega_o^2 \right)\ln\left(1+(2ar)^2\right)/(8a^2)$ can be interpreted as a {\it potential energy}. Therefore, its minima set by
$\delta V/\delta r=\left(\Omega_o^2 -\Omega^2 \right) r/\left(1+(2ar)^2\right)=0$ and $\delta^2 V/\delta r^2=\left(\Omega_o^2 -\Omega^2 \right) \cdot \left((1-(2ar)^2)\right)/\left(1+(2ar)^2\right)^2\geq 0$, is the full equilibrium position. It confirms that for $\Omega <\Omega_o$ the stable equilibrium position is found at $r=0$ whereas for $\Omega >\Omega_o$ the sphere is pushed up to infinity.
\subsubsection{Frictional case}
We consider now the motion of a real sphere. Therefore, we must introduce the rolling friction in the dynamical equation of motion. For simplicity, we forbid the sliding. Projected on the radial coordinate, the equation is now:
\begin{equation}
\ddot{ r}= r\Omega^2 -\left( r\Omega^2 \sin(\alpha)+g\cos(\alpha)\right)\sin(\alpha) -sign(\dot{r})\mu_r\left(r\Omega^2 \sin(\alpha)+g\cos(\alpha)\right)\cos(\alpha)
\label{StabFrict1}
\end{equation}
\noindent
where $sign(\dot{r})=1$ for $\dot{r}\geq 1$ and $sign(\dot{r})=-1$ for $\dot{r}<1$.
Following the same steps than for the frictionless case, this equation can be integrated to get:
\begin{equation}
\frac{d}{d\tau}\left(\frac {y ^2}{2}\right)=\delta E_o+ \left(x^2 -1\right)\frac{\ln\left(1+y^2\right)}{2} -sign(\dot{y})\mu_r\left[x^2 y+(1-x^2)\arctan(y)\right]
\label{StabFric2}
\end{equation}
\noindent
where, as previously, we use the dimensionless variable: $x=\Omega/\Omega_o$ and $y=2ar$. We introduce also the dimensionless time $\tau=1/\Omega_o$. Once again, one can write equation (\ref{StabFric2}) as: $ \delta E_o=K(\dot{y})+V(y,x)$ with 
$V(y,x)=-\left(x^2 -1\right)\ln\left(1+y^2\right)/2+sign(\dot{y})\mu_r\left[x^2 y+(1-x^2)\arctan(y)\right]$. Moreover the centrifugal force pushes away the sphere and the gravity attracts it to the center (cf figure\ref{FigDevCent}). Therefore, one can reasonably assume that the sign of $\dot{r}$ is given by the sign of the tangential force applied on the particle, thus by the sign of $\Omega^2 r \cos(\alpha)-g\sin(\alpha)$. Hence 
$sign(\dot{y})\equiv sign(x^2-1)$. It will be the surely case if no velocity impulse is induced by the initial perturbation or if we consider the particle has lost the memory of the direction of the initial velocity perturbations. The equilibrium condition $\delta V(x,y)/\delta y=0$ reduces to:
\begin{equation}
\mu_r x^2 y^2-|x^2-1|+\mu_r=0
\label{polymur}
\end{equation}
\noindent 
which has exactly the same roots than eq. (\ref{frictionpolyn}) excepted that the rolling friction, $\mu_r$, replaces the static friction, $\mu_o$, with $\mu_r\leq\mu_o$ in general.
Figure \ref{req}, shows that smaller is $\mu_o$, smaller is the gap enclosed by the root of equation \ref{frictionpolyn}. Thus, the solutions of equation (\ref{polymur}) are enclosed in the static equilibrium area computed in the previous section. More precisely, taking into account the sign of $(x^2-1)$, one can show that the stable equilibrium set by the minimum of $V(x,y)$ are given by $y^-=(|x^2-1|-\sqrt{|x^2-1|^2-4\mu_r^2 x^2})/(2\mu_r x^2)$ for $\Omega<\Omega_o$ (i.e $x>1$) and $y^+=(|x^2-1|+\sqrt{|x^2-1|^2-4\mu_r^2 x^2})/(2\mu_r x^2)$ for $\Omega>\Omega_o$. In this last case, the particle escapes from the cell, since stable equilibrium radius is larger than the cell radius (see figure \ref{req}). The equilibrium position deduced from static considerations can be then destabilized under an external perturbation. However, an injection of kinetic momentum is necessary to destabilize the static position. This perturbation has to overcome the onset imposed by the static friction. This threshold is easier to overcome near the boundary of the static area. In any case, it seems interesting to compare the rolling friction involved in the stability analysis with the static friction.
\section{Transient regime and rolling friction}
\begin{figure}
\centering
\includegraphics[width= 1.0\columnwidth,angle=0]{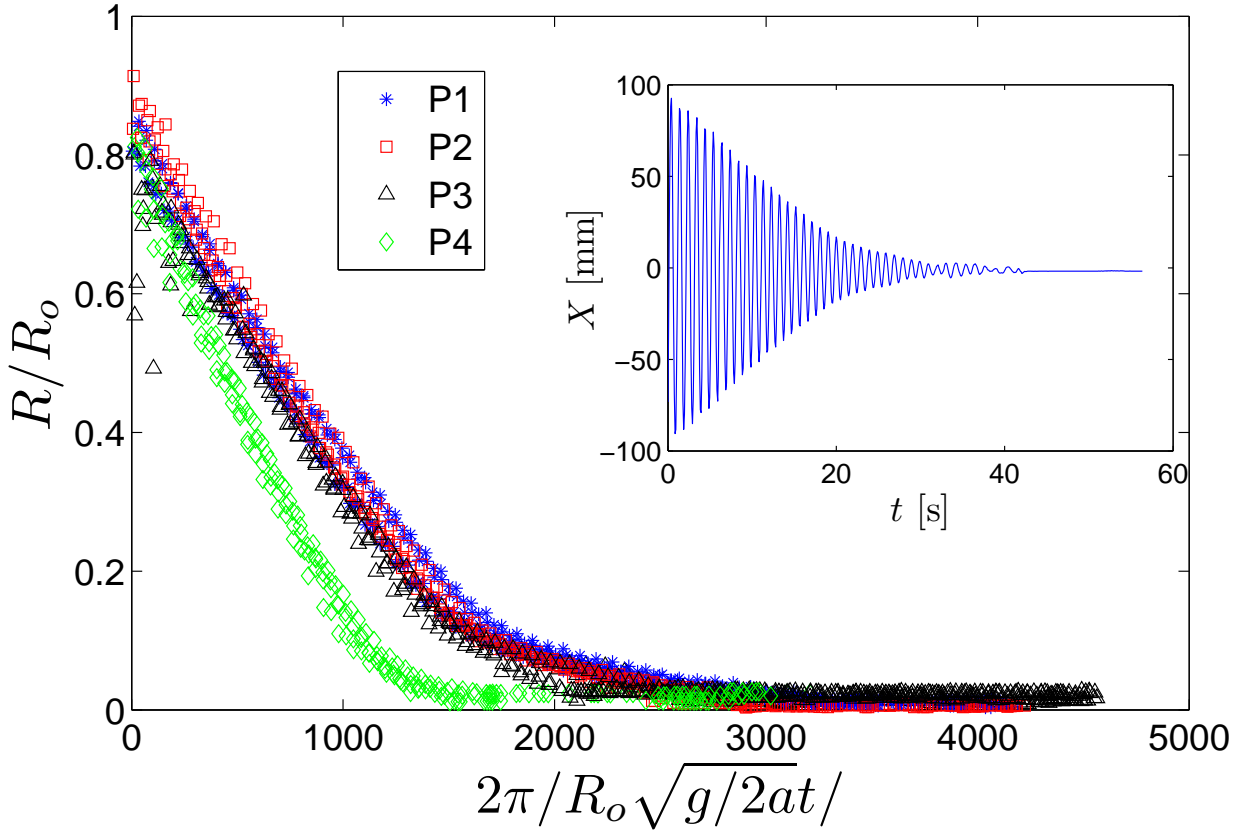}
\caption{Main Panel: Decay of the amplitude of the oscillatory motion (shown in the inset) of the spheres rolling freely on the motionless parabola. Four successive launching are shown for each spheres: the 6mm diameter stainless steel spheres, P1 (blue asterix), P2 (red square), the glass sphere of 6 mm diameter, P3 (black up-triangle) and the 3 mm diameter stainless steel sphere P4 (green diamond). The time unit is chosen to get access to the rolling friction by the measure of the slope of the initial linear decay. Inset: The x--component of the oscillatory damping motion of the sphere P1 in physical unit. }
\label{Decay}
\end{figure}

To reach it static position or to escape from a metastable position, the particle has to roll and thus must experience rolling friction. Therefore in order to describe the dynamic of our particles, we must estimate this rolling friction in our cell. Moreover there are still open questions concerning the rolling friction. Actually several mechanisms can be evocated to explain the dissipation during the rolling \cite{Johnson}. In addition to the fundamental visco--plasticity of the materials in contact \cite{Poschel,Johnson}, one can also raise the loss of energy by micro--sliding \cite{Johnson}, emission of phonon and sound (the rolling can be heart in the experiment), micro--collisions, electrostatic interaction or viscous effect of the surrounding air etc....
The combination of these effects makes difficult the general predictions for the rolling friction properties.
For instance, if it generally is accepted that the slide friction depends strongly on the load and very weakly on the velocity, this question is less clear for the rolling friction. The dependence of the rolling friction coefficient with the sphere velocity is still debated. Contradictory results exist and it seems to be highly sensitive to the experimental procedure\cite{Witters,Domenech2,Shaw,Xu}. This is why we must measure the rolling friction directly in our own experimental setup. In order to simplify the problem, without loss of generality, we consider only the case $\Omega=0$. Therefore the centrifugal and Coriolis forces vanish. The static equilibrium position of the sphere is therefore $r_{eq}=0$. We let a particle falling freely from the largest radius of the parabola. Then we follow it until its equilibrium position. For each particle, we repeat 4 launching at different positions of the cell circumference.
The x-position of the particle for one launch of the particle P1, is shown in the inset of figure \ref{Decay}. It presents a clear damped oscillation. The oscillation period is about $0.99\pm0.00$s for each particles and therefore quite compatible with the intrinsic period: $2\pi/\Omega_o=0.95$. Let us assuming that the radial position of the sphere can be write $r(t)=R(t)|\sin(\Omega_ot)|$ with $R(t)$ the decreasing envelop of the oscillation. This envelop is shown in the main panel of the figure \ref{Decay} for all the spheres and all the launching. At large amplitude, this decay is not exponential. Such an exponential decay would have been the signature of a viscous damping, i.e. of a rolling friction coefficient proportional to the velocity. Actually, it appears to be linear until small amplitude value where the surface imperfections become sensitive (especially for the small sphere). Starting from the cell outer diameter $R_o$, one expects that $R(t)=R_o(1-t/\tau)$ at the beginning. The linear decay rate $\tau$ seems to be similar for the large spheres whatever are the material. It is larger for the smaller sphere.
To understand such decay, one can try to estimate the energy lost induced by friction during an oscillating period. We can consider that the friction is mainly proportional to the load at the beginning of the decaying motion. Indeed, it seems not associat with an exponential viscous damping at that stage. Moreover, one can assume that the decay rate per period is small. The loss of kinetic energy is therefore $\Delta K\sim m\overline{v}\delta v$. It must be balanced by the work of the friction force during this period $\Delta W=-\mu_r m g \overline{v} T$, where $\overline{v}$ is the mean velocity amplitude over period, $T=2\pi/\Omega_o$, and $\delta v$ is the velocity change during $T$. Hence $\delta v\sim\delta R/\delta t\sim \mu_r g T$ and $R(t)\sim R_o\left (1-\mu_r\frac{2\pi}{R_o}\sqrt{\frac{g}{2a}}\cdot t\right )$. With the choice of the unit of time used in figure \ref{Decay}, the slope of the initial linear decay gives a direct access to the rolling friction. The value of the rolling friction deduced from figure \ref{Decay} are given in Tab. 2.
The fit are made for $\frac{2\pi}{R_o}\sqrt{\frac{g}{2a}}\cdot t\leq 1000$. Error bars are estimated from the different initial launching conditions. Finally, one must check that the characteristic damping time $\tau$ is larger the period $T$ to justify the expansion made for a small decay rate by period. This is indeed the case since: $\tau/T=R_o/(\mu_r g)\sim 25$.
The friction coefficients reported in Tab. 2 are 100 times smaller than the static friction. Hence, the fully stable positions are nearly mingled with the asymptotical case of frictionless particles. This very low rolling friction is of order of magnitude of what it is referred as the {\it railroad steel wheel on steel rail} and twice smaller than in ref.\cite{Domenech2,Wikifriction} but we may use a harder anodized aluminum support. Moreover it is still two order of magnitude larger than the theoretical prediction of ref.\cite{Poschel}. However, note that the simple balance proposed above, does not explain why the friction coefficient is larger for a smaller sphere made of the same material. Such a particle size dependency of the rolling friction has been already reported \cite{Domenech2}. A possible explanation implies its smaller weight which facilitates the jumps of the sphere on asperities hence a higher loss of energy by inelastic collisions. The air effects are also more sensitive where the diameter is smaller. A systematic study of the friction with the sphere diameter is out of the scope of our work. 
\section{Conclusions}
The work presented here mixed several concepts of classical mechanics in a single experiment: friction, centrifugal force, support reaction, static balance and stability analysis. The use of parabolic support allows analytical computations. These predictions can be directly compared to experimental measurements. Moreover the setup allows direct estimates of both the static and rolling friction of different spherical materials. We show that the static position of the sphere is compatible with the lower bound of the static friction measured on an inclined plan. It may be interesting to try to interpret the small discrepancy between the two measurement methods. Is it related to the two different angles observed for the start and the stop of sand avalanches \cite{DaerrDouady}? Moreover, the damping oscillations of a freely rolling sphere give an estimate of the rolling friction. We show that it depends poorly of the velocity and that it is hundred times smaller than the static friction
In the presented results, the friction seems to be almost the same for glass and stainless steel spheres which are both hard materials. However the experiment can be also performed with much more soft materials like wood or soft plastic balls. The dissipation mechanism should be increased, but the precision on the sphere shape will be lower. The plastic deformations could also have some impacts. The effect of the sphere radius on the static and rolling friction has to be probed farther, by using larger spheres for instance. To probe properly the air effect, it is necessary to perform a secondary vacuum into the rotating cell. This would increase the complexity of our rotating device. Finally, in order to go deeper in the granular matter world, it could be interesting to study the collective behavior of a set of spheres on this device. The particles will experience some additive dissipation due to the mutual friction and the inelastic collisions. We should be able to answer to the following questions: What will be the area occupied by the spheres? How is it related to $r_{eq}$ ? What will be the structure adopted by the grains? What will be the grains dynamics? How all these properties are related to the dissipative processes induced between grains? Are we going to observe some kind of segregation induced by friction or shape disparity? etc... .
%
%
\begin{acknowledgments}
We gratefully acknowledge Vincent Padilla who built up the parabola. This work benefits of the fruitful discussions with Anne Tanguy, Jean-Christophe Geminard and Antoine Naert. This work has been supported by Triangle de la physique contract 2011--075T -- COMIGS2D
\end{acknowledgments}

\end{document}